\documentclass[aps,prd,twocolumn,nofootinbib,showpacs,superscriptaddress]{revtex4-1}

\usepackage{color}
\usepackage{bm}
\usepackage{amsfonts,amssymb}
\expandafter\let\csname equation*\endcsname\relax
  \expandafter\let\csname endequation*\endcsname\relax
\usepackage{amsmath}
\usepackage{graphicx,graphics}
\usepackage{enumerate}
\usepackage{subfigure}
\usepackage{appendix}
\usepackage{siunitx}
\usepackage{mathrsfs}   

\usepackage[latin1]{inputenc}
\usepackage{hyperref}

\def\be{\begin{equation}}
\def\ee{\end{equation}}

\def\mbf{\mathbf}

\definecolor{Gruen}{rgb}{.322,.537,.035}

\begin{document}
\title{Rotating black holes in a draining bathtub: superradiant scattering of gravity waves}
\author{Maur\'icio Richartz}
\email{mauricio.richartz@ufabc.edu.br}
\affiliation{Centro de Matem\'atica, Computa\c{c}\~ao e Cogni\c{c}\~ao, Universidade Federal do ABC (UFABC), 09210-170 Santo Andr\'e, SP, Brazil.}
\author{Angus Prain}
\email{aprain@ubishops.ca}
\affiliation{Physics Department and STAR Research Cluster, Bishop's University, 2600 College St.,
Sherbrooke, Queebec, Canada J1M 1Z7}
\author{Stefano Liberati}
\email{liberati@sissa.it}
\affiliation{SISSA - International School for Advanced Studies
via Bonomea 265, 34136 Trieste, Italy, and
INFN, Sezione di Trieste}
\author{Silke Weinfurtner}
\email{silke.weinfurtner@nottingham.co.uk}
\affiliation{School of Mathematical Sciences, University of Nottingham, University Park, Nottingham, NG7 2RD, UK}

\begin{abstract}

In a draining rotating fluid flow background, surface perturbations behave as a scalar field on a rotating effective black hole spacetime. We propose a new model for the background flow which takes into account the varying depth of the water. Numerical integration of the associated Klein-Gordon equation using accessible experimental parameters shows that gravity waves in an appropriate frequency range are amplified through the mechanism of superradiance. Our numerical results suggest that the observation of this phenomenon in a common fluid mechanical system is within experimental reach. Unlike the case of wave scattering around Kerr black holes, which depends only on one dimensionless background parameter (the ratio $a/M$ between the specific angular momentum and the mass of the black hole), our system depends on two dimensionless background parameters, namely the normalized angular velocity and surface gravity at the effective black hole horizon. 
\end{abstract}


\maketitle
%
%
\section{Introduction and Motivation \label{sec:Motivation}}
%
%

Analogue models of gravity have, for some time now, been an excellent arena within which to improve our theoretical understanding of several crucial phenomena at the boundary of gravity and quantum field theory. The first model, proposed by Unruh in 1981, was based on the fact that sound waves propagating on an inviscid and irrotational fluid flow satisfy a Klein--Gordon (KG) equation in an effective curved background \cite{unruh}. If the velocity of the fluid exceeds the velocity of sound at some closed surface, a dumb hole -- the analogue of a black hole horizon for sound waves -- forms. Since Unruh's seminal paper, the propagation of perturbations in many other physical systems have been shown to be analogous to that of fields on a curved spacetime (see \cite{Barcelo:2005fc} for a survey and review). 
                        
One particularly relevant phenomenon that can be simulated in an analogue spacetime is the evaporation of black holes by Hawking radiation. Indeed, one of the first theoretical goals of analogue gravity was to investigate the dependence of Hawking radiation on the arbitrarily high frequencies used in its original derivation (the trans-Planckian problem). Calculations with a modified dispersion relation at high frequencies gave support to the reality of the evaporation process \cite{jacobson, unruh_2, brout} and simultaneously paved the way for a possible experimental observation of the analogue Hawking process in tabletop experiments.

Experimental research on analogue systems started only very recently. The first analogues of an event horizon were constructed in the laboratory using gravity waves on water \cite{unruh_3} and ultrashort pulses in optical fibers \cite{philbin}. In 2010, the classical analogue of the stimulated emission by a white hole was detected in the laboratory for the first time \cite{weinfurtner, weinfurtner2}. Since a white hole is the time reversal of a black hole, this result attests to the generality of the Hawking radiation process. At the same time, some first hints of Hawking radiation in optical systems were reported~\cite{belgiorno,faccio}. More recently, claims to link an instability arising in critical superfluid flows and Hawking radiation appeared~\cite{stein,nguyen}.

Another potential target for analogue gravity is the experimental observation of superradiance \cite{basak,basak2}, a phenomenon in which incident waves are amplified after being reflected by a special kind of scattering potential. Superradiance~\cite{beken, mauricio} was first studied by Zel'Dovich for electromagnetic waves incident on a conductive rotating cylinder \cite{zeldovich}, but also pertains to black holes~\cite{staro1, staro2} and analogue black holes~\cite{basak,basak2, berti, lepe, ednilton, disper}. See Ref.~\cite{super_review} for a recent review on the subject.

In a combination of theory and numerical simulations, our primary scientific goal in this work is to show the existence and estimate the magnitude of the superradiant amplification in a realistic scattering scenario of gravity waves on a common draining water vortex (i.e.~a realistic axisymmetric draining fluid). Put simply, we wish to understand whether the common draining `bathtub' vortex is a suitable system for the onset of superradiant scattering and whether any amplification is experimentally observable. 

The paper is structured as follows. In the second section we present in detail our description of the background flow for a draining vortex, moving beyond the standard description found in the literature.  Following this, in Sec.~\ref{analogue_metric} we describe free surface perturbations on the background profile and show that they satisfy a KG equation of motion for an auxiliary metric, the analogue metric for gravity waves. The development is largely pedagogical up to this point. With the basic machinery in place we then specify to the black hole-like analogue geometries of draining rotating vortices (Sec.~\ref{sec_solve}) and discuss the limits of our approximations (Sec.~\ref{Sec:limits}). The existence of superradiant scattering is derived in Sec.~\ref{sec4}. In Sec.~\ref{Sec:rescale}, we show that, by re-scaling the parameters and variables, our problem can be completely described by only two dimensionless background parameters and two dimensionless wave parameters. This situation is then compared with the scattering around a Kerr black hole. The results of our numerical simulations are presented in Sec.~\ref{numerics}, where we obtain the spectrum of reflection coefficients for several background flows. In particular, we show that the observation of superradiance in a common fluid mechanical system is within experimental reach. Finally, Sec.~\ref{Sec:finalremarks} is devoted to our final remarks and conclusions.

\section{Background flow} \label{back_flow}

The system considered in this work is the propagation of shallow water gravity waves on the air-water interface of an open channel flow over a flat bottom. In general, an open-channel incompressible perfect fluid configuration is described by a free-surface function $h(t,\mbf{x})$ and a flow velocity vector field $\mbf{v}(t,\mbf{x})$. These variables must satisfy the standard continuity and Euler equations,
\begin{align}
\nabla\cdot\mbf{v}&=0,\label{E:Cont}\\
\frac{\partial \mbf{v}}{\partial t}+\left(\mbf{v}\cdot \nabla\right)\mbf{v}&=-\frac{\nabla P}{\rho}-g\mbf{\hat{z}}, \label{E:Euler}
\end{align}
where $\mbf{\hat{z}}$ is the direction of action of the restoring gravitational force, $g$ is the gravitational acceleration, $P$ is the pressure and $\rho$ the density function. We will be primarily concerned with the behaviour of the fluid at the free surface (since this is where gravity waves propagate). Therefore, let us introduce a system of coordinates adapted to this free surface, $\mbf{x}=(z,\mbf{x}_\parallel)$, where $z$ is the coordinate measuring vertical displacements (defined with respect to the direction of action of the gravitational force) and $\mbf{x}_\parallel$ are coordinates orthogonal to $z$.  

The hydrodynamic equations above are subject to a set of boundary conditions appropriate to the physical system in question, namely,

\begin{enumerate}[A.]
\item The normal flow velocity must vanish at the bottom of the channel, i.e.~$\left. v_{z} \right| _{z=0} = 0$; \label{enum_1}
\item The rate of change in the height of the fluid must be equal to the vertical velocity of the fluid at the surface\label{enum_2}
\begin{align} \label{bound1}
\left. v_z \right|_{z=h} &= \left. \frac{dh}{dt} \right| _{z=h} = \frac{\partial h}{\partial t} +  \left(\left. \mathbf{v}_{\parallel}\right| _{z=h} \cdot \nabla_{\parallel}\right) h; 
\end{align}
\item The pressure must be continuous at the air-water interface. \label{enum_3}

\end{enumerate}

It is our goal in this preliminary section to determine the equations for the free surface $h$ and the velocity field $\mbf{v}$ in a draining bathtub vortex configuration. Firstly, let us assume that the flow is irrotational, i.e.~$\nabla\times\mbf{v}=0$. This assumption reduces the vector field $\mbf{v}$ to one scalar degree of freedom $\psi$, the so called velocity potential defined by $\mbf{v}(t,\mbf{x})=\nabla \psi(t,\mbf{x})$. In terms of this potential, the continuity equation \eqref{E:Cont} reduces to Laplace's equation $\nabla^2 \psi = 0$.  

The analogy with gravity arises in the regime of shallow water perturbations~\cite{unruh_ralf,rouss}. Therefore, it is sensible to expand the field $\psi$ in powers of the vertical displacement $z$. Such a series expansion is very common in hydrodynamics and was first introduced by Lagrange~\cite{darrigol}. Indeed, using Laplace's equation together with the boundary condition~\eqref{enum_1}, one can write the velocity potential at any point as~\cite{bouss}
\be
\psi\left(t,\mathbf{x_{\parallel}},z\right)= \sum_{n=0}^{\infty} \frac{(-1)^n z^{2n}}{(2n)!} \nabla_{\parallel}^{2n} \psi_{0}\left(t,\mathbf{x_{\parallel}}\right), \label{E:psi}
\ee
where $\psi_0\left(t,\mathbf{x_{\parallel}}\right)=\left. \psi \left(t,\mathbf{x_{\parallel}},z\right) \right|_{z=0}$ is the velocity potential at the flat bottom.
 
Finally, assuming a stationary and axisymmetric flow, it is convenient to adopt cylindrical coordinates $\mbf{x}=(z, r,\phi)$. In such a setup, the free surface function $h$ depends only on $r$ and the irrotational flow condition $\nabla\times \mbf{v}=0$ implies that $v_\phi={B}/{r}$, where $B$ is a constant. Then the scalar function $\psi_0$ takes the form
$\psi_0(r,\phi)= \xi(r)+B\phi$, where $\xi(r)$ is an unknown function of $r$. 

Additionally, the continuity and Euler equations combine with the boundary conditions to give
\be
\frac{1}{h} \int _0 ^{h} v_r(r,z) dz \equiv
\sum_{n=0}^{\infty} \frac{(-1)^n h^{2n}}{(2n+1)!} \frac{\partial}{\partial r} \mathcal{D}^{2n} \xi(r) = - \frac{C}{r h}, \label{E:A}
\ee
and
\begin{align}
\left. v_r(r,z) \right|_{z=h(r)}\equiv
\sum_{n=0}^{\infty} \frac{(-1)^n h^{2n}}{(2n)!} & \frac{\partial}{\partial r} \mathcal{D} ^{2n} \xi(r) \nonumber
\\  =\frac{-1}{ \sqrt{1 + h'^2}} & \sqrt{ 2 g \left(h_{\infty} - h \right) - \frac{B^2}{r^2}}, \label{E:B}
\end{align}
where $v_r(r,z)= \partial_r \psi$ is the radial flow velocity, $C$ is a constant of integration and  prime denotes differentiation with respect to $r$. The differential operator $\mathcal{D}^2$ is the radial part of the Laplacian, i.e.~
\be
\mathcal{D}^{2}=\frac{1}{r}\frac{\partial}{\partial r}r\frac{\partial}{\partial r}.
\ee
%

In summary the background configuration is described completely by the two univariate functions $h(r)$ and $\xi(r)$, and the constants $B$ and $C$ (which are subject to the equation of motion and boundary conditions discussed above).  Physically, we expect that in the limit of large radius the flow velocities all vanish ($\mbf{v}\rightarrow 0$) and that the free surface approaches a constant ($h\rightarrow h_\infty$).
Note that the only approximations used so far are those related to the assumption of a perfect incompressible fluid in an 
axisymmetric and irrotational flow.

\section{Gravity waves and their effective geometry} \label{analogue_metric}

When the free surface of an open channel flow is perturbed, gravity acts as a restoring force, creating oscillations around the background flow. The mathematical description of these oscillations, called gravity waves, is given in terms of linear perturbations $\delta \psi_{0} \ll \psi_{0}$ and $ \delta h \ll h$ of the background quantities $\psi_{0}$ and $h$\footnote{Another possibility would be to perturb the background flow around $\left. \psi \left(t,\mathbf{x_{\parallel}},z\right) \right|_{z=h(r)}$ instead of perturbing around $\left. \psi \left(t,\mathbf{x_{\parallel}},z\right) \right|_{z=0} = \psi_0$ as we have done.}. Indeed, by linearizing both Euler's equation \eqref{E:Euler} (evaluated at the free surface and written in terms of $\psi_0$) and the boundary condition \ref{enum_2}, one obtains the following pair of coupled equations,
\begin{widetext}
\begin{equation} 
\frac{\partial \delta h}{\partial t} + \nabla_{\parallel} \cdot \left( \left. \mathbf{v}_{\parallel}\right|_{z=h} \delta h \right) 
\label{pert3} +  \nabla_{\parallel} \cdot \left(  \sum_{n=0}^{\infty} \frac{(-1)^n h^{2n+1}}{(2n+1)!} \nabla_{\parallel}^{2n+1} \delta \psi_{0} \right)= \frac{\partial \delta h}{\partial t} + \nabla_{\parallel} \cdot \left( \left. \mathbf{v}_{\parallel}\right|_{z=h} \delta h \right) 
 +  \nabla_{\parallel} \cdot \left( h \nabla_{\parallel} \delta \psi_{0} \right) =  0, 
\end{equation}
and
\begin{equation} \label{pert4}
\left(\frac{\partial}{\partial t} + \left. \mathbf{v_{\parallel}}\right|_{z=h} \cdot \nabla_{\parallel}  \right) \left( \sum_{n=0}^{\infty} \frac{(-1)^n h^{2n}}{(2n)!} \nabla_{\parallel}^{2n} \delta \psi_{0}   \right)
    =  \left(\frac{\partial}{\partial t} + \left. \mathbf{v_{\parallel}}\right|_{z=h} \cdot \nabla_{\parallel}  \right)\delta \psi_{0} 
= -\tilde g \delta h,
\end{equation}
\end{widetext}
where $\mathbf{v_{\parallel}} = \nabla_{\parallel} \psi$ and $\tilde g$ is given by
\begin{equation} \label{tildeg}
\tilde g = g + \left(\frac{\partial}{\partial t} + \left.\mathbf{v}_{\parallel}\right|_{z=h} \cdot \nabla _{\parallel}\right)^2 h.
\end{equation}

Note that, in order to obtain the first equality in Eqs.~\eqref{pert3} and \eqref{pert4}, we have assumed that the height of the fluid $h$ is much smaller than the wavelength $\lambda$ of the perturbations. When acting on such long wavelengths, one has $\nabla _{\parallel}^2 = O \left(1/\lambda ^2 \right)$ and $h/\lambda \ll 1$ \cite{unruh_ralf,rouss}, so that we need to keep only the lowest order terms in the sums above (we shall return to this assumption in Sec.~\ref{Sec:limits}). 

Eliminating $\delta h$, we obtain a single second order differential equation for $\delta \psi_{0}$ which can be written as a KG equation for a scalar field in a curved Lorenzian spacetime,
\begin{equation} 
\label{KG} 
\frac{1}{\sqrt{-\mathbf{g}}}\partial_{\mu}(\sqrt{-\mathbf{g}}g^{\mu \nu} \partial_{\nu}\delta \psi)=0,
\end{equation}
where $g_{\mu \nu}$ is an effective metric given by
\begin{equation} \label{metric}
g_{\mu \nu} = \left(\frac{h}{\tilde g}\right) \left( \begin{array}{cc}
-\tilde g h +\left. v_{\parallel}^2 \right|_{z=h} & -\left. \mathbf{v_{\parallel}}\right|_{z=h} \\
-\left.\mathbf{v_{\parallel}}\right|_{z=h} & \mathbf{I} _{2 \times 2}
\end{array} \right),
\end{equation}
and $\mathbf{g}=$det$(g_{\mu\nu})$.

The above expression for the analogue geometry implies that the velocity of gravity waves is given by 
\begin{equation}
c_{\mathrm{gw}}(r)=\sqrt{\tilde{g}h(r)}, \label{E:w_speed}
\end{equation}
generalizing the standard result~\cite{unruh_ralf,rouss} through the modified (and non-constant) `gravitational acceleration' $\tilde{g}$. Depending on the relation between the wave speed and the flow velocity, this metric can describe the analogues of both an event horizon and an ergoregion experienced by gravity waves in this system, as we shall discuss below.

\section{Analogue black hole description} \label{sec_solve}

One important advantage of working in the analogue gravity framework is that quantities which require very subtle and technical definitions in general relativity acquire simple and intuitive definitions in terms of the fluid parameters of the analogue system. For example, assuming again stationarity and axisymmetry, we can say that the analogue of an ergoregion is the region where the surface fluid velocity exceeds the propagation speed for gravity waves, i.e.~
\be
\left. v_{\parallel}^2 \right|_{z=h}=\left. v_\phi^2 \right|_{z=h} + \left. v_r^2 \right|_{z=h} > c_{\mathrm{gw}}^2, \quad (\text{ergoregion})\label{E:ergo_region}
\ee
and that the event horizon is the surface on which the radial velocity alone is equal to the propagation speed, i.e.~
\be
\left. v_r^2 \right|_{z=h}=c_{\mathrm{gw}}^2 \quad (\text{event horizon}). \label{E:event_horizon}
\ee

One important characterizing quantity associated with horizons is the surface gravity $\kappa$, which fixes a scale for processes that occur at or near the horizon.  In general relativity this is an excruciatingly subtle parameter to define~\cite{Cropp:2013zxi} whereas for our analogue fluid horizon it is simply expressed~\cite{Visser:1997ux} as
\be
\kappa_\mathrm{H}=\frac{1}{2} \frac{d}{dr} \left. \left( c_{\mathrm{gw}}^2 - v_r^2  \right) \right| _{\text{horizon}} \, . \label{E:surface_gravity}
\ee

In principle, by imposing suitable boundary conditions, one can exactly solve Eqs.~\eqref{E:A} and \eqref{E:B} to determine the free surface $h(r)$ and the function $\xi(r)$, which can then be used to determine the other relevant background quantities. The problem, however, is that Eqs.~\eqref{E:A} and \eqref{E:B}, being infinite order differential equations, are very difficult to solve. Instead, we make the vastly simplifying assumption of slowly varying $v_r(r,z)$ as a function of $z$ (that there is no separation of flow), implying 
\be
 \frac{1}{h} \int _0 ^{h} v_r(r,z) dz \approx \left. v_r(r,z) \right|_{z=h(r)}.
\ee
In conjunction with~\eqref{E:A} this leads to
\be
\left. v_r(r,z) \right|_{z=h(r)} = - \frac{A h_{\infty}}{r h},  \label{E:radial_3} 
\ee
where $A$ is a constant obtained by redefining the constant $C$~\footnote{Note that expression \eqref{E:radial_3} is insensitive to the precise dependence of $v_r$ on $z$, the same result following from the assumption of any monomial dependence $v_r\propto z^p$ or, more generally, any polynomial in $z$, $v_r=\sum_p^Na_p z^p$, with coefficients $a_p$ such that $a_p\propto h^{N-p}$, lending credibility to the general validity of expression \eqref{E:radial_3}. Such polynomial expressions are justified by the fact that, in realistic flows, we expect that $\left. v_r(r,z) \right|_{z=0}=0$ due to the no-slip boundary condition at the bottom.}. With the expression above, equation \eqref{E:B} becomes a first order differential equation for $h(r)$,
\be
\frac{A^2 h_\infty ^2}{r^2 h^2} \left(1 + h'^2\right)= 2 g \left(h_{\infty} - h \right) - \frac{B^2}{r^2}, \label{E:radial_4}
\ee
while the expression \eqref{tildeg} for $\tilde{g}$ becomes 
\begin{align} 
\tilde g= g + \frac{A^2 h_{\infty}^2}{r^2h^2} \left(h'' - \frac{h'}{r} - \frac{h'^2}{h} \right).   \label{tildeg2}
\end{align}

Finally, we make the assumption that $h'(r)^2 \ll 1$, reducing~\eqref{E:radial_4} to the cubic polynomial
\be
 h^3 + h^2\left[\frac{B^2}{2g r^2} - h_{\infty} \right]  + \frac{A^2h_{\infty}^2}{2gr^2} = 0, \label{E:poly}
\ee
which admits the physical solution
\begin{align} 
h(r)=\frac{1}{3} \left( h_{\infty} - \frac{B^2}{2g r^2} \right) \left[1+2 \cos \left( \frac{\theta}{3} \right) \right],
\label{height} 
\end{align}
with
\begin{align}
\theta=\cos ^{-1}\left[ 1 - \frac{3^3}{g} \left(\frac{Ah_{\infty}}{2r}\right)^2 \left(  h_{\infty} - \frac{B^2}{2g r^2} \right)^{-3} \right]. 
\end{align}
Substituting \eqref{E:radial_3} and \eqref{tildeg2} into \eqref{E:event_horizon}, gives
\be
\frac{A^2 h_{\infty}^2}{r^2h^2} = gh + \frac{A^2 h_{\infty}^2}{r^2h} \left(h'' - \frac{h'}{r} - \frac{h'^2}{h} \right)  \label{E:event}
\ee
which (after substituting the explicit form \eqref{height} of $h(r)$ derived above), can be solved for $r$ in order to locate the event horizon of the system.

\section{Limits of our approximations} \label{Sec:limits}

We have employed two main approximations for the background flow profile in order to render the problem more tractable. Firstly we assume a slowly varying $v_r(r,z)$ as a function of $z$ to obtain relation \eqref{E:radial_3} for the radial flow profile. Secondly we assume $h'^2\ll1$ resulting in a closed form expression for the free surface. 
We expect both of these approximations to break down near the central core of the vortex where vorticity and viscosity begin to play a role. In experimental work~\cite{andersen2003anatomy,andersen2006bathtub} on the structure of the bathtub vortex it has been shown that the flow is not irrotational near the core of such vortices and that the flow velocities are regulated there (avoiding the unphysical $1/r$ flow profile singularities in $v_r$ and $v_\phi$ which we use here). We point out however that the free surface solution of \eqref{E:poly} for draining fluids has been considered before in a different context by Refs.~\cite{1977WP,mang}, where it has shown remarkable agreement with experimental data. 

On the other hand, we employ a single approximation in the description of the gravity waves themselves by neglecting higher order terms in the sums of \eqref{pert3} and \eqref{pert4}. This is the shallow water approximation which results in linearly dispersive gravity waves. This kind of `linear dispersion' approximation is ubiquitous in analogue gravity since, beyond the realm of linear dispersion, the description of linear perturbations in terms of a single effective metric is lacking.
In fact, assuming $h=h_\infty=\text{constant}$ and keeping all terms in the sums \eqref{pert3} and \eqref{pert4}, leads exactly to the usual gravity wave dispersion relation
\be
\omega^2=gk\tanh(kh_\infty), \label{E:dispersion}
\ee
where $k$ and $\omega$ are, respectively, the wavenumber and the frequency of the wave. In such a case, the shallow water approximation is characterized by the condition $kh_\infty \ll 1$, under which \eqref{E:dispersion} becomes the linear dispersion relation
\be
\omega^2=gh_\infty k ^2.
\ee
Notice that the condition $kh_\infty \ll 1$, in view of the linear dispersion above, becomes $\omega \ll \sqrt{g/h_{\infty}}$. For the sake of concreteness, let us define linear waves as those satisfying
\be \label{E:omegadisp} 
\omega < \omega_{\text{disp}} = \sigma \, \sqrt{\frac{g}{h_\infty}},
\ee
where $\sigma$ is a constant and $\omega_\text{disp}$ sets the dispersive scale\footnote{The factor $\sigma$ is based on a subjective measure of where the dispersion curve is visibly linear in character. In our numerical work we have chosen to use the conservative value $\sigma=0.3$ .}. 

It is important to keep in mind that, crucially, the phenomena themselves which are of experimental interest in the analogue gravity community (in our case superradiance \cite{disper} but also in the case of the Hawking process \cite{jacobson,PhysRevD.48.728, Corley:1996ar,PhysRevD.57.6280,unruh_2, brout,Coutant:2011in}) are relatively robust to modifications to the linear dispersion relation and do not require per se a universal metric structure (effective or not). This suggests that these phenomena are generic, to which the spacetime relativistic or analogue linear-dispersive versions are special cases.

\section{Superradiance} \label{sec4}

We have already shown that the propagation of gravity waves obeys a KG equation in an analogue spacetime [see Eq.~\eqref{KG}]. This analogue spacetime is determined by the background open channel flow described in section~\ref{back_flow}. Furthermore, because of the axisymmetry of the system, the KG equation~\eqref{KG} is separable. The ansatz $\delta \psi_0(t,r,\phi)=R(r)e^{im\phi}e^{-i\omega t}$, where $m$ is the azimuthal number of the wave, transforms equation~\eqref{KG} into   
\begin{equation} \label{radialeqn}
\frac{d^2R}{dr^2}  + P(r) \frac{dR}{dr} + Q(r) \, R=0,
\end{equation}
where the coefficients $P(r)$ and $Q(r)$ are given by
\begin{widetext}
\begin{align}
P(r) &= \frac{d}{dr}  \log\left[\frac{r}{\tilde{g}} \left(\tilde{g}h- \left. v_r ^ 2 \right|_{z=h} \right) \right] + 2 \, i \, \frac{ \left. v_r \right|_{z=h}}{\tilde{g}h -\left. v_r^ 2 \right|_{z=h} } \left( \omega - m\frac{B}{r^2} \right),\label{Eq:P_general}\\
Q(r) &= \frac{1}{\tilde{g}h - \left. v_r ^2 \right|_{z=h} }\left[  \left( \omega - m\frac{B}{r^2} \right)^2 - m^2\frac{\tilde{g}h}{r^2} \right] + \frac{i}{r}  \frac{\tilde g}{\tilde{g}h-\left. v_r ^2 \right|_{z=h} }\frac{d}{dr}  \left[\frac{r}{\tilde{g}} \left. v_r \right|_{z=h}  \left( \omega - m\frac{B}{r^2} \right) \right].\label{Eq:Q_general}
\end{align}
\end{widetext}

As explained above, the background radial flow velocity $\left. v_r \right|_{z=h}$ is determined  by~\eqref{E:radial_3}, $\tilde g$ is calculated from~\eqref{tildeg2}, and the flow depth $h(r)$ is given by~\eqref{height}. Similarly to the procedure used when studying real black hole perturbations, in order to analyze the asymptotic behavior of these hydrodynamic perturbations, it is useful to define a tortoise-like coordinate $r_*$ according to    
\begin{equation} \label{tortoise}
\frac{dr_{*}}{dr}= \Delta(r) = \tilde{g}h \left( \tilde{g}h - \left. v_r ^2 \right|_{z=h} \right)^{-1}.
\end{equation}
While the original radial coordinate ranges from $r=r_\mathrm{H}$ near the horizon to $r=\infty$ at spatial infinity, this tortoise coordinate ranges from $-\infty$ to $\infty$. The precise relation between these coordinates, in the asymptotic limits, is given by
\begin{equation} \label{tortlimit}
r_* \rightarrow
\left\{
\begin{array}{ll}
\frac{c_\mathrm{H} ^2}{2 \kappa _\mathrm{H}} \log (r-r_\mathrm{H}) , \ \ \ & r \rightarrow r_\mathrm{H},
\\
r,  & r \rightarrow \infty,
\end{array}
\right. 
\end{equation}
where $\kappa_\mathrm{H}$ is the analogue of the surface gravity [see \eqref{E:surface_gravity}]
and $c_\mathrm{H}=c_{\mathrm{gw}}(r_\mathrm{H})$ is the propagation speed of the waves at the horizon. 

Defining a new radial function $H(r)$ by
\begin{equation}
H(r) = \Delta ^{1/2} \exp \left[\frac{1}{2} \int^{r} P(u)du \right] R(r),
\end{equation}
\\
we eliminate the first order term in Eq.~\eqref{radialeqn}, obtaining
\begin{equation} \label{waveq}
\frac{d^2 H}{d{r_{*}}^{2}} +  V[r(r_*)] \, H = 0,
\end{equation}
where the potential $V$ is given in terms of $r=r(r_*)$ by the expression
\begin{widetext}
\begin{align} 
V(r)  &= \frac{1}{\tilde{g}h}\left( \omega - \frac{m \, B}{r^2} \right)^2 - \frac{m^2}{\Delta r^2} + \frac{1}{2}\left(\frac{(h r)'}{h r} \right) \frac{\Delta '}{\Delta^3}- \frac{1}{\Delta^2}\left[ \frac{1}{4} \left(\frac{(h r)'}{h r} \right)^2 + \frac{1}{2}\left(\frac{(h r)'}{h r} \right)' \right].  \label{effk2}
\end{align}
Although complicated to look at, \eqref{effk2} possesses the simple $r_*$ independent asymptotics 
\begin{equation}
V(r)\stackrel{r\rightarrow r_\mathrm{H}}{\longrightarrow} \frac{1}{c_\mathrm{H}^2} \left(\omega-\frac{mB}{r_\mathrm{H}^2}\right)^2 \quad \text{ and } \quad
V(r)\stackrel{r\rightarrow +\infty}{\longrightarrow} \frac{\omega^2}{gh_\infty},
\end{equation}
allowing us to write the solution of the wave equation corresponding to the scattering of an incoming wave from $r=+\infty$ as 
\begin{equation} \label{hsolution1}
H(r_*) =
\left\{
\begin{array}{ll}
\alpha_\text{in} e^{-i \frac{\omega}{c_{\infty}} r_*} + \alpha_\text{out} e^{+i \frac{\omega}{c_{\infty}} r_*},  & r_* \rightarrow +\infty \,\,(r \rightarrow \infty),
\\
\alpha_\text{tr} e^{ -\frac{i}{c_\mathrm{H}}\left( \omega - \frac{mB}{r_\mathrm{H} ^2} \right)  r_*}, \ \ \ & r_* \rightarrow -\infty \,\,(r \rightarrow r_\mathrm{H}),
\end{array}
\right.
\end{equation}
\end{widetext}
where $c_{\infty} = \sqrt{gh_{\infty}}$ is the wave speed far away from the black hole and $\alpha_\text{in}$, $\alpha_\text{out}$ and $\alpha_\text{tr}$ are constants. In obtaining the expression above, we have used the boundary condition that no signal can escape from inside the analogue event horizon. 

The constants $\alpha_\text{in}$, $\alpha_\text{out}$ and $\alpha_\text{tr}$ are not all independent. Using the fact that the Wronskian between two solutions of Eq.~\eqref{waveq} is independent of $r_*$ and that the complex conjugate of~\eqref{hsolution1} is also a solution of the wave equation, we conclude that 
\begin{equation} \label{conserva1}
\frac{\omega}{c_{\infty}}\left(1 - \left|\frac{\alpha_{out}}{\alpha_{in}}\right|^2 \right) = \frac{1}{c_\mathrm{H}}\left( \omega - \frac{mB}{r_\mathrm{H} ^2} \right) \left|\frac{\alpha_{tr}}{\alpha_{in}}\right|^2,
\end{equation}
from which the following reflection $\mathcal{R}$ and transmission $\mathcal{T}$ coefficients, satisfying $\mathcal{R} + \mathcal{T}=1$, can be defined:
\begin{equation}\label{r_t}
\mathcal{R}=\left|\frac{\alpha_{out}}{\alpha_{in}}\right|^2, \qquad \mathcal{T}=\frac{c_{\infty}}{\omega c_\mathrm{H}}\left( \omega - \frac{mB}{r_\mathrm{H} ^2} \right) \left|\frac{\alpha_{tr}}{\alpha_{in}}\right|^2.
\end{equation}
Superradiance occurs when the norm of the reflected wave is greater than the norm of the incident wave, that is, when the reflection coefficient is greater than 1 (equivalently, when the transmission coefficient is negative). We conclude, therefore, that superradiance occurs whenever
\begin{equation}
0< \omega < \frac{mB}{r_\mathrm{H} ^2}. \label{supcon}
\end{equation}

This derivation of superradiance is similar, but not identical, to the usual derivation for internal pressure waves~\cite{basak}. Typical internal pressure waves are characterized by a constant density $\rho$, a constant wave speed $c$, and a radial flow velocity with simple radial dependence $v_r = -A/r$. Gravity waves in our paper, on the other hand, are characterized by a variable fluid depth $h(r)$, a variable wave speed $c(r) = \sqrt{\tilde g h}$ and a radial flow velocity $v_r$ given by equation \eqref{E:radial_3}. If $h(r)$ was constant ($\tilde g$ would reduce to $g$ in such a case), then $v_r$ and $c$ would reduce to the simple expressions valid for pressure waves, and the two systems would be basically the same. This is exactly what is studied in Ref.\cite{berti}. Note that, in order to have a constant $h(r)$ in an open channel flow, Refs.\cite{berti,unruh_ralf} assume a non-flat bottom (together with some restrictions on the slope of such a bottom). In our analysis, however, we have a flat bottom and a non-constant $h(r)$. Therefore, one cannot simply use the expressions for internal waves derived in Ref.~\cite{basak}, like the authors of Ref.~\cite{berti} were able to do. We have shown that, no matter what the radial dependence of h(r) is, the phenomenon of superradiance will always occur.

\section{Re-scaling and comparison with Kerr black holes} \label{Sec:rescale}

So far we have worked in terms of  dimensionful parameters and dimensionful variables. In this section, we show that the variable background parameters $A,B$, $h_\infty$ together with $g$ can be combined into exactly two independent variable dimensionless parameters which, together with the parameters $m$ and $\omega$ associated with the wave, are sufficient to completely describe the scattering problem.

\begin{figure*}[!htb]
\begin{center}
\centering
\includegraphics[scale=1.2]{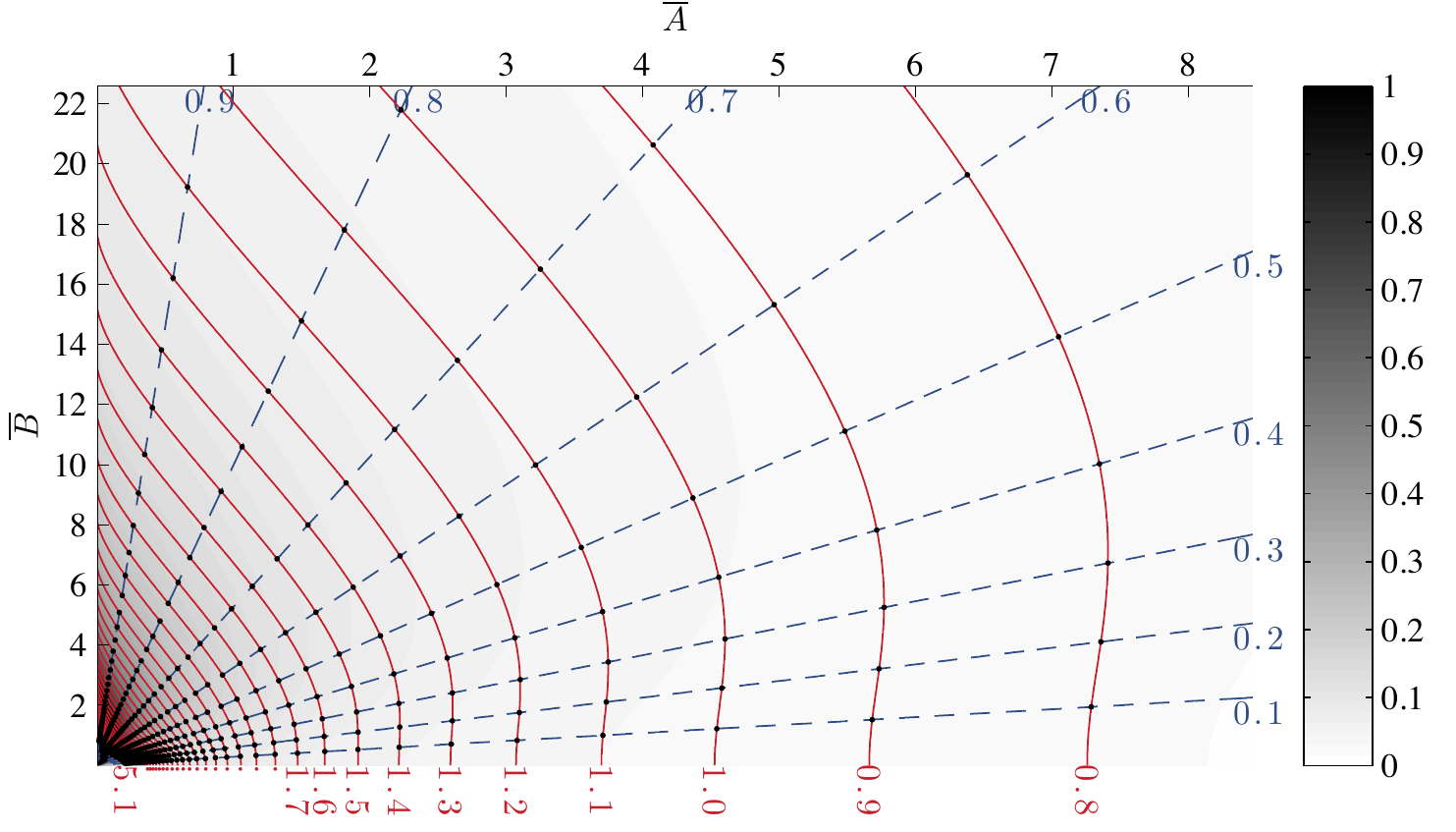}
\caption{(Colors online.) Lines of constant $\overline{v}_\phi (r_\mathrm{H})$ (dashed blue lines) and $\overline{\kappa}(r_\mathrm{H})$ (solid red lines) are plotted over the cartesian $\overline{A},\overline{B}$ plane for $\overline{A}=[0,8]$ and $\overline{B}=[0,22]$. The greyscale represents the value of $h'(\overline{r}_\mathrm{H})$ for the background configuration at each point in parameter space (we have artificially cut off the grayscale at $1$ to improve visualization).  The intersections of the $\overline{v}_\phi (r_\mathrm{H})$ and $\overline{\kappa}(r_\mathrm{H})$ contour lines (indicated by small black dots) are used in our numerical simulations of superradiant scattering.}
\label{F:fig1}%
\end{center}
\end{figure*}

We start by re-scaling the cylindrical coordinates $r$ and $z$ of our system using the length scale $h_{\infty}$ according to $\overline{r}=r/h_\infty$ and $\overline{z}=z/h_\infty$. This naturally defines a dimensionless free surface function $\overline{h}=h/h_\infty$. In terms of these dimensionless variables, the cubic polynomial \eqref{E:poly}, whose solution is the free surface function $h(r)$, can be written as 
\be
\overline{h}^3+\overline{h}^2\left(\frac{ \overline{B}^2}{\overline{r}^2}-1\right)+\frac{\overline{A}^2}{ \overline{r}^2}=0, \label{E:h}
\ee
where $\overline{A}$ and $\overline{B}$ are dimensionless and given by
\begin{align}
\overline{A}&=\frac{A}{h_\infty\sqrt{2gh_\infty}},\\
\overline{B}&=\frac{B}{h_\infty\sqrt{2gh_\infty}}. \label{E:tilde}
\end{align}
As we can see, the scaled free surface function $\overline{h}$ is completely characterized by the two parameters $\overline{A}$ and $\overline{B}$. Similarly, in terms of the dimensionless quantities, expression \eqref{tildeg2} for $\tilde g$ can be written as
\be
\tilde{g}=g\left[1+\frac{2\overline{A}^2}{\overline{r}^2\overline{h}^2}\left(\overline{h}_{,\overline{r} \overline{r}}-\frac{\overline{h}_{,\overline{r}}}{\overline{r}}-\frac{\overline{h}_{,\overline{r}}^2}{\overline{h}}\right)\right],
\ee
so that $\overline{g}=\tilde{g}/g$ is also dimensionless. Equation \eqref{E:event} defining the location of event horizon, on the other hand, can be written as 
\be
\overline{h}+\frac{2\overline{A}^2}{\overline{r}^2\overline{h}}\left(\overline{h}_{,\overline{r} \overline{r}}-\frac{\overline{h}_{,\overline{r}}}{\overline{r}}-\frac{\overline{h}_{,\overline{r}}^2}{\overline{h}}-\frac{1}{\overline{h}}\right)=0. \label{E:horiz}
\ee
Therefore, since $\overline{h}$ depends only on $\overline{A}$ and $\overline{B}$, both $\overline{g}$ and the dimensionless location of the event horizon $\overline{r}_\mathrm{H}=r_\mathrm{H}/h_{\infty}$ will also be completely determined by $\overline{A}$ and $\overline{B}$.

Finally, the wave equation \eqref{waveq} around our analogue black hole can be re-scaled as
\be \label{E:waveq_ad}
\left[\frac{ d^2}{d\overline{r}_*^2}+\overline{V}(\overline{r}_*)\right]H(\overline{r}_*)=0,
\ee
where $\overline{r}_*=r_*/h_{\infty}$ is the dimensionless tortoise coordinate, 
\begin{align}
\overline{V}=\frac{2}{\overline{g}\overline{h}}\left(\frac{\sigma \overline{\omega}}{\sqrt{2}}- \frac{m\overline{B}}{\overline{r}^2}\right)^2 + \frac{1}{2}\left(\frac{(\overline{h} \overline{r}),_{\overline{r}}}{\overline{h} \overline{r}} \right) \frac{\Delta,_{\overline{r}}}{\Delta^3} \nonumber \\ 
-\frac{m}{{\Delta}\overline{r}^2} - \frac{1}{\Delta^2}\left[ \frac{1}{4} \left(\frac{(\overline{h} \overline{r}),_{\overline{r}}}{\overline{h} \overline{r}} \right)^2 + \frac{1}{2}\left(\frac{(\overline{h} \overline{r}),_{\overline{r}}}{\overline{h} \overline{r}} \right),_{\overline{r}} \right]
\end{align}
is the dimensionless equivalent to $V$ and 
\begin{align}
\overline{\omega}&= \frac{\omega}{\omega_\text{disp}}= \frac{\omega}{\sigma}\sqrt{\frac{h_\infty}{g}}
\end{align}
is the dimensionless frequency defined by scaling $\omega$ with \eqref{E:omegadisp}. Note also that the dimensionless function $\Delta$ can be written in terms of the dimensionless quantities as 
\be
\Delta= \tilde{g}h \left( \tilde{g}h - \left. v_r ^2 \right|_{z=h} \right)^{-1} = \frac{\overline{h}}{\overline{h}-\frac{\overline{A}}{\overline{r}^2\overline{h}^2}}.
\ee

Therefore the scattering of waves governed by \eqref{E:waveq_ad} depends exclusively on the parameters $\overline{A}$, $\overline{B}$ of the background flow and on the parameters $\overline{\omega}$ and $m$ which characterize the waves. Contrast this result with the superradiant scattering of scalar waves around a Kerr black hole, which is described by two dimensionful background parameters $a$ and $M$ (respectively, the specific angular momentum and the mass of the black hole) and three wave parameters: $\ell$, $m$ and $\omega$ (respectively, the azimuthal number, the orbital number, and the frequency of the wave). After re-scaling, the background will be described by a single dimensionless parameter $a/M$ (in units of $G=c=1$). The largest possible amplification occurs when the black hole approaches extremality, i.e.~$a/M \rightarrow 1$. Indeed, it has been show that the maximum possible amplification is 0.3\% for scalar waves~\cite{p_teuko1}, 4.4\% for electromagnetic waves and 138\% for gravitational waves~\cite{p_teuko2}.

Note that $a/M=2 \Omega r_+$, where $\Omega$ is the angular velocity of the Kerr black hole and $r_+$ is the location of its event horizon.  This result naturally generalizes to analogue black holes and suggests that the quantity $\Omega r_\mathrm{H} = v_{\phi}|_{r=r_\mathrm{H}} = B/r_\mathrm{H}$ might be important. Remarkably, similarly to the extremality condition for real black holes, there is also a limit for the quantity $B/r_\mathrm{H}$ in analogue black holes. From Bernoulli's equation, it is straightforward to show that $v_{\phi}|_{r=r_\mathrm{H}} = B/r_\mathrm{H} < \sqrt{2gh_{\infty}}$. Physically this is nothing more than conservation of energy: a fluid packet far from the vortex has energy given by the gravitational potential alone $E/\rho= gh_\infty$; at the horizon, the maximum possible rotational flow velocity is achieved when all this energy is converted to rotational kinetic energy, or $g h_\infty=v_\phi^2/2$, from whence the bound follows. 

   It is important to point out that, from a mathematical point of view, any two (independent) dimensionless combinations of $A,B,h_\infty$ and $g$ are sufficient to describe the background flow and no special role is played by $\overline{A}$ and $\overline{B}$. In view of this, we will work with two distinct dimensionless parameters which allow for a natural and immediate comparison with the Kerr black hole scattering, namely the normalized rotational flow velocity evaluated at the event horizon $\overline{v}_\phi(r_\mathrm{H})$ and the scaled surface gravity $\overline{\kappa}(r_\mathrm{H})=\kappa_\mathrm{H}/g$:
\be
\overline{v}_\phi(r_\mathrm{H}) =\left. \frac{v_\phi}{\sqrt{2gh_\infty}}\right|_{r_\mathrm{H}} =\frac{\overline{B}}{\overline{r}_\mathrm{H}},  \label{E:Dim1} 
\ee
\be
\overline{\kappa}(r_\mathrm{H})=\frac{1}{2}\left.\frac{d}{d\overline{r}}\left[\overline{h} +\frac{2\overline{A}^2}{\overline{r}^2\overline{h}}\left(\overline{h}_{,\overline{r} \overline{r}}-\frac{\overline{h}_{,\overline{r}}}{\overline{r}}-\frac{\overline{h}_{,\overline{r}}^2}{\overline{h}}-\frac{1}{\overline{h}}\right)\right]\right|_{\overline{r}_\mathrm{H}} \label{E:kappa1},
\ee
where the definition \eqref{E:surface_gravity}, together with $v_\phi=B/r$, have been used. 

As we can see, both $\overline{\kappa}(r_\mathrm{H})$ and $\overline{v}_\phi(r_\mathrm{H})$ are functions determined uniquely by the two parameters $\overline{A}$ and $\overline{B}$. One can therefore, write all the equations in this section in terms of $\overline{v}_\phi(r_\mathrm{H})$ and $\overline{\kappa}(r_\mathrm{H})$, instead of $\overline{A}$ and $\overline{B}$. However, since the definitions of $\overline{\kappa}(r_\mathrm{H})$ and $\overline{v}_\phi(r_\mathrm{H})$ involve the event horizon coordinate $r_\mathrm{H}$ which is the solution to a difficult non-linear equation, the transformation between ($\overline{\kappa}(r_\mathrm{H})$, $\overline{v}_\phi(r_\mathrm{H})$) and ($\overline{A}$, $\overline{B}$) is non-trivial and needs to be done numerically. In Fig.\ref{F:fig1} we show this transformation explicitly by performing a parameter search in the $(\overline{A}$, $\overline{B})$ space to find lines of constant $\overline{v}_\phi(r_\mathrm{H})$ and $\overline{\kappa}(r_\mathrm{H})$. We have also indicated a grey scale for the value of $h'(r_\mathrm{H})$ at each point of the parameter space, which represents a measure of the validity of the approximation $h'(r)^2\ll 1 $ used in our model of the background flow.

\section{Numerical results}\label{numerics}

\subsection{Methodology}

Let us first explain how the reflection and transmission coefficients $\mathcal{R}$ and $\mathcal{T}$ are obtained numerically. We have already shown how the background flow and the free surface profile are completely determined by the three dimensionful parameters $A$, $B$ and $h_{\infty}$. For a given choice of these parameters one can in principle solve equation~\eqref{waveq} with the boundary conditions given in~\eqref{hsolution1}. For numerical reasons we found it easier to solve equation~\eqref{radialeqn} directly instead of \eqref{waveq}. By doing so, we avoid having to invert the equation relating the usual $r$ coordinate and the tortoise coordinate $r_*$. The asymptotic form of the fields, given by \eqref{hsolution1}, written in terms of the function $R(r)$ become
\begin{widetext}
\begin{equation} \label{rsolution1}
R(r) =
\left\{
\begin{array}{ll}
\beta_\text{in} r^{-\frac{1}{2}+i \frac{A\omega}{c_{\infty}^2}} e^{-i \frac{\omega}{c_{\infty}} r} + \beta_\text{out} r^{-\frac{1}{2}+i \frac{A\omega}{c_{\infty}^2}} e^{+i \frac{\omega}{c_{\infty}} r}, \quad  & r \rightarrow +\infty,
\\
\beta_\text{tr}[1+K_1(r-r_\mathrm{H})], \qquad & r \rightarrow r_\mathrm{H},
\end{array}
\right.
\end{equation}
where
\be
K_1= -\frac{ \left(\omega - m \frac{B}{r_\mathrm{H}^2} \right)^2 - \frac{m^2c_\mathrm{H}^2}{r_\mathrm{H}^2} + i \frac{c_\mathrm{H}^2}{r_\mathrm{H} h_\mathrm{H}} \frac{d}{dr}\left.\left(\frac{r v_r|_{z=h}}{\tilde g}  \left(\omega - m \frac{B}{r_\mathrm{H}^2} \right) \right)\right|_{r=r_\mathrm{H}} }{2 \kappa_\mathrm{H} \left(1 - i \frac{ch}{\kappa_\mathrm{H}} \left(\omega - m \frac{B}{r_\mathrm{H}^2} \right) \right)}.
\ee
\end{widetext}
The asymptotic behaviour near the analogue black hole horizon naturally translates into the following boundary condition for Eq.~\eqref{radialeqn} at the point $r_\text{min}=r_\mathrm{H} + \epsilon$ (where $\epsilon \ll 1$): $R(r_\text{min})=\beta_\text{tr}=1$ and $R'(r_\text{min})=\beta_\text{tr} K_1 = K_1$. After solving Eq.~\eqref{radialeqn} numerically for a given frequency $\omega$, we are able to obtain $R(r_\text{max})$ and $R'(r_\text{max})$, where the point $r_\text{max}\gg r_\text{min}$ is located sufficiently far away from the event horizon. With the help of the asymptotic expansions above, we are then able to use $R(r_\text{max})$ and $R'(r_\text{max})$ to determine $\beta_\text{in}$ and $\beta_\text{out}$. 

The reflection and transmission coefficients, given by equation~\eqref{r_t}, when written in terms of $\beta$'s instead of $\alpha$'s, become (recall that $\beta_\text{tr}=1$): 
\begin{equation}\label{r_t_2}
\mathcal{R}=\left|\frac{\beta_\text{out}}{\beta_\text{in}}\right|^2, \qquad \mathcal{T}=\frac{4c_{\infty}h_\mathrm{H} r_\mathrm{H}}{\omega c_\mathrm{H} h_{\infty}r_\text{max}}\left( \omega - \frac{mB}{r_\mathrm{H} ^2} \right) \left|\frac{1}{\alpha_\text{in}}\right|^2
\end{equation}       
where $h_\mathrm{H}$ is the depth of the water at the horizon. Using the numerically determined values for $\beta_\text{in}$ and $\beta_\text{out}$, we are able to obtain $\mathcal{R}$ and  $\mathcal{T}$ for a given pair of the parameters $\omega$ and $m$. 

In our simulations, we have used $h_{\infty}=\SI{4}{\centi \meter}$, $A=[0,3000] \ \SI{}{\centi \meter ^2/s}$ and $B=[0, 8000] \ \SI{}{\centi \meter ^2/s}$, so that the whole $\overline{A},\overline{B}$ parameter space shown in Fig.~\ref{F:fig1} is covered. This way we can easily convert the used parameters to the corresponding parameters $\overline{\kappa}(r_\mathrm{H})$ and $\overline{v}_\phi(r_\mathrm{H})$. Furthermore, we have used $\epsilon =\SI{10}{^{-8}\centi \meter}$ and $r_\text{max}=\SI{10000}{\centi \meter}$. For a given value of the parameter $m$, we repeat the procedure described above for several frequencies in the range $0 < \omega < mB/r_\mathrm{H}^2$, thus obtaining the spectrum of reflection coefficients in the superradiant regime. Finally, by locating the maximum of each curve, we are able to determine the maximum possible amplification $\mathcal{R}_\text{max}$ of the spectrum and the corresponding frequency $\omega_\text{peak}$.

\subsection{Results and discussion}

%
\begin{figure*}
\centering
\subfigure[$\,$ Varying $\overline{v}_\phi(r_\mathrm{H})$ for $\overline{\kappa}(r_\mathrm{H})=1.2$. \label{WaterProfileVnormp4_10cm}]
{\includegraphics{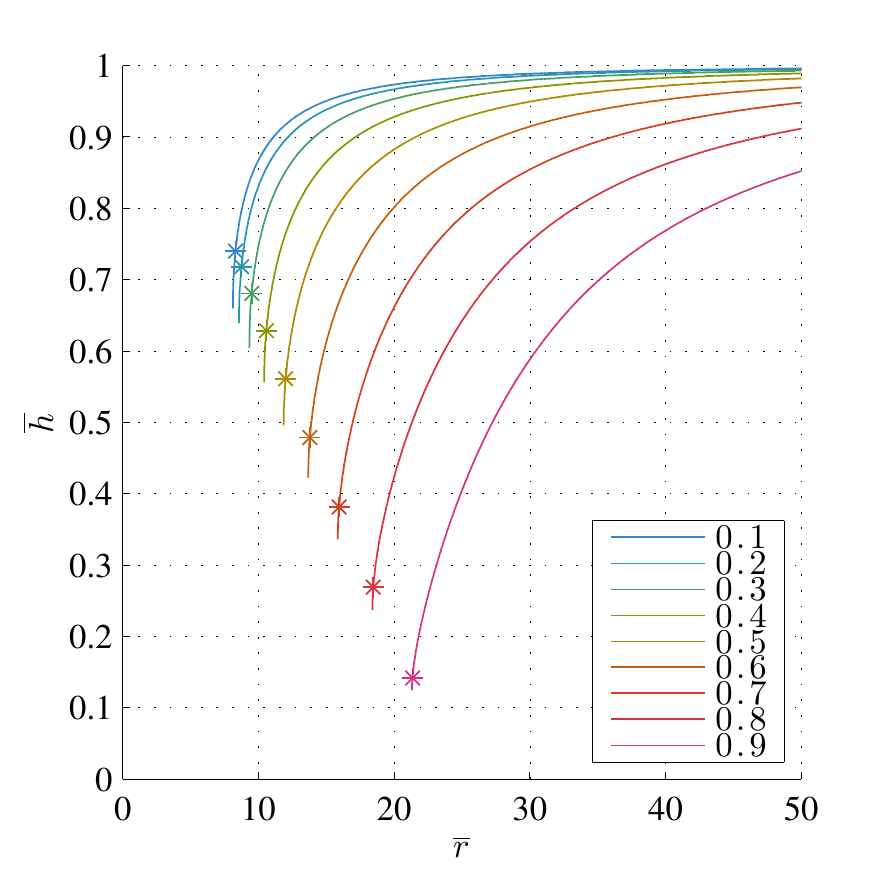}}
\subfigure[$\,$ Varying $\overline{\kappa}(r_\mathrm{H})$ for $\overline{v}_\phi(r_\mathrm{H})=0.4$.\label{WaterProfileVnormp4_4cm}]{\includegraphics{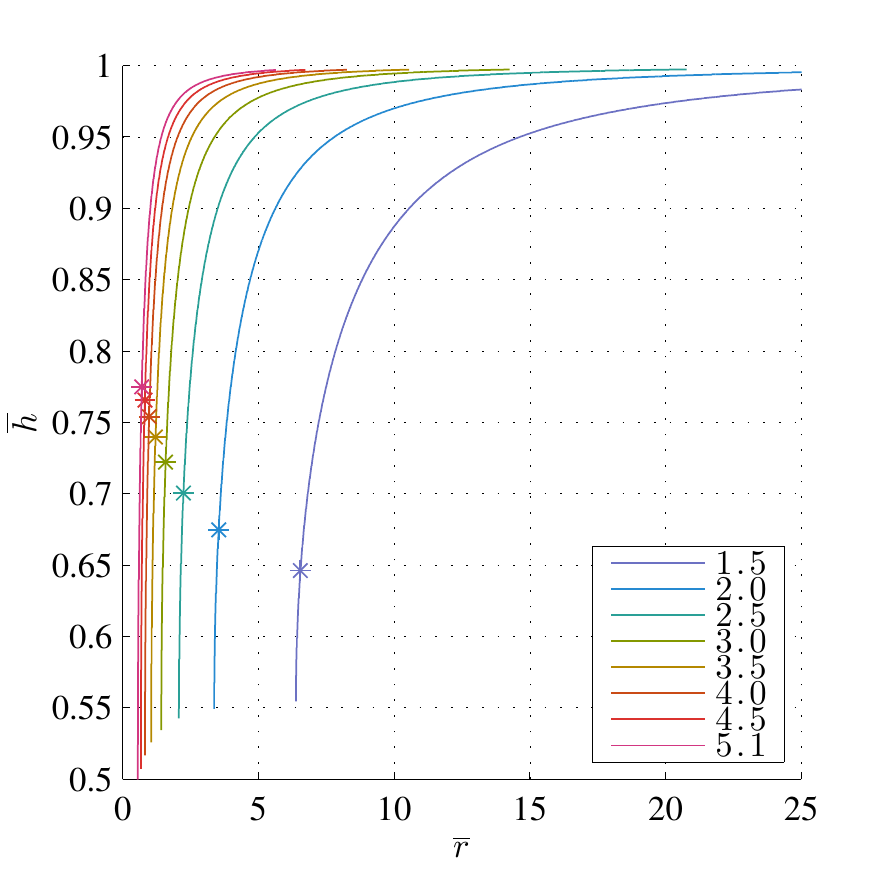}}
\caption{(Colors online.) The water profiles (solid lines) and the corresponding location of the event horizon (asterisks), as computed in Eqs.~\eqref{E:h} and \eqref{E:horiz}, for several values of the (a) normalized angular velocity and the (b) normalized surface gravity at the effective black hole horizon.}\label{F:fig4}%
\end{figure*}
%

In this section we present the numerical results. As explained above, the dimensionless free surface of the fluid can be completely determined by the parameters $\overline{v}_\phi(r_\mathrm{H})$ and $\overline{\kappa}(r_\mathrm{H})$. With the help of Fig.~\ref{F:fig1} and Eq.~\eqref{height} we plot $\overline{h}(\overline{r})$ in Fig.~\ref{F:fig4} for different values of these background parameters. For each curve, the location of the event horizon, found using \eqref{E:horiz}, is indicated by an asterisk. Note that $\overline{h}(\overline{r})$ is always a strictly increasing function of $\overline{r}$, so that the slope of the water surface $\overline{h}'(\overline{r})$ is always a decreasing function. Consequently, in the region of interest for the scattering process itself, which ranges from the effective event horizon to infinity, the point which has the largest possible slope is exactly the event horizon. Therefore, we can use the value of $\overline{h}'(\overline{r}_\mathrm{H})$ as a measure of the validity the assumption $h'^2 \ll 1$ used in our model. In this sense, Fig.~\ref{F:fig1} shows that lower surface gravity profiles are better approximated by our model, below a value of about $\overline{\kappa}(r_\mathrm{H})\simeq 3$. The approximation seems to be rather insensitive to the value of $\overline{v}_\phi(r_\mathrm{H})$, although more critical flows ($\overline{v}_\phi(r_\mathrm{H})$ approaching 1) have slight larger values of $\overline{h}'(\overline{r}_\mathrm{H})$ when compared to less critical flows.

\begin{figure*}[!htb]
\centering
\subfigure[$\,$ Varying $\overline{v}_\phi( r_\mathrm{H})$ for $\overline{\kappa}( r_\mathrm{H})=1.2$.\label{SpectraConstS2100_4cm}]{\includegraphics{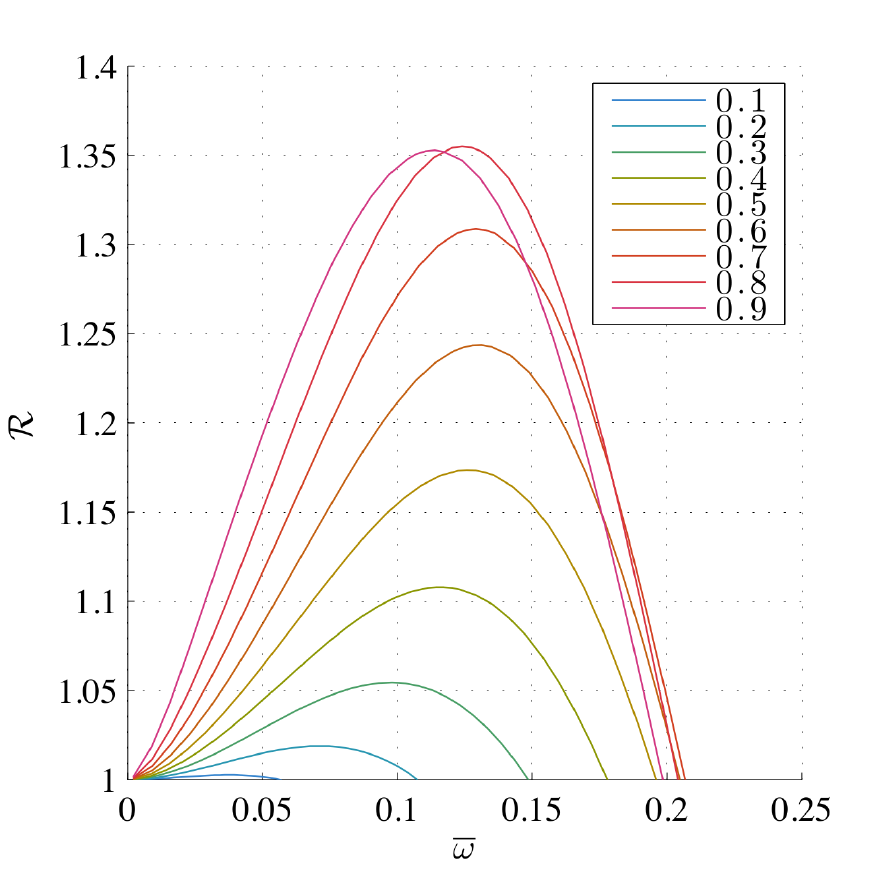}}
\subfigure[$\,$ Varying $\overline{\kappa}( r_\mathrm{H})$ for $\overline{v}_\phi( r_\mathrm{H})=0.4$.\label{SpectraConstVnormp4_4cm}]{\includegraphics{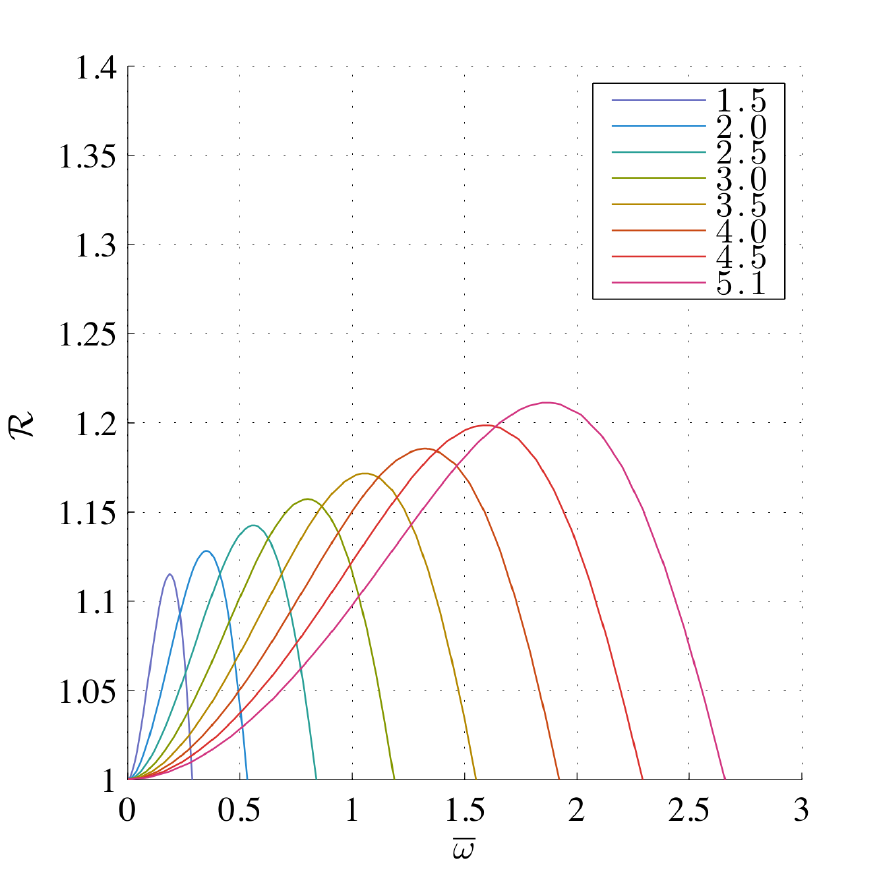}}
\caption{(Colors online.) The spectra of reflection coefficients in the superradiant regime for several values of the (a) normalized angular velocity and the (b)  normalized surface gravity at the effective black hole horizon.}\label{F:fig3}
\end{figure*}

\begin{figure*}[!htb]
\centering
\subfigure[$\,$ Maximum reflection coefficient $\mathcal{R}_\text{max}$.\label{RmaxVersusKappaPaper_4cm}]{\includegraphics{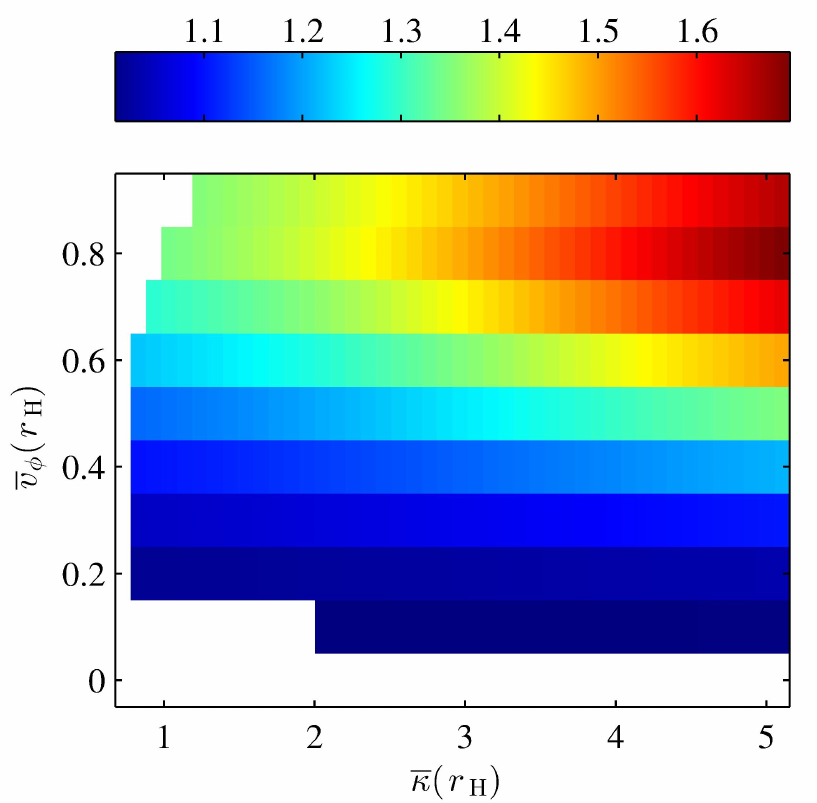}}
\subfigure[$\,$ Scaled frequency $\overline{\omega}_\text{peak}$ corresponding to the maximum reflection coefficient. \label{OmegaMaxOverOmegaDispersionPaper_4cm}]{\includegraphics{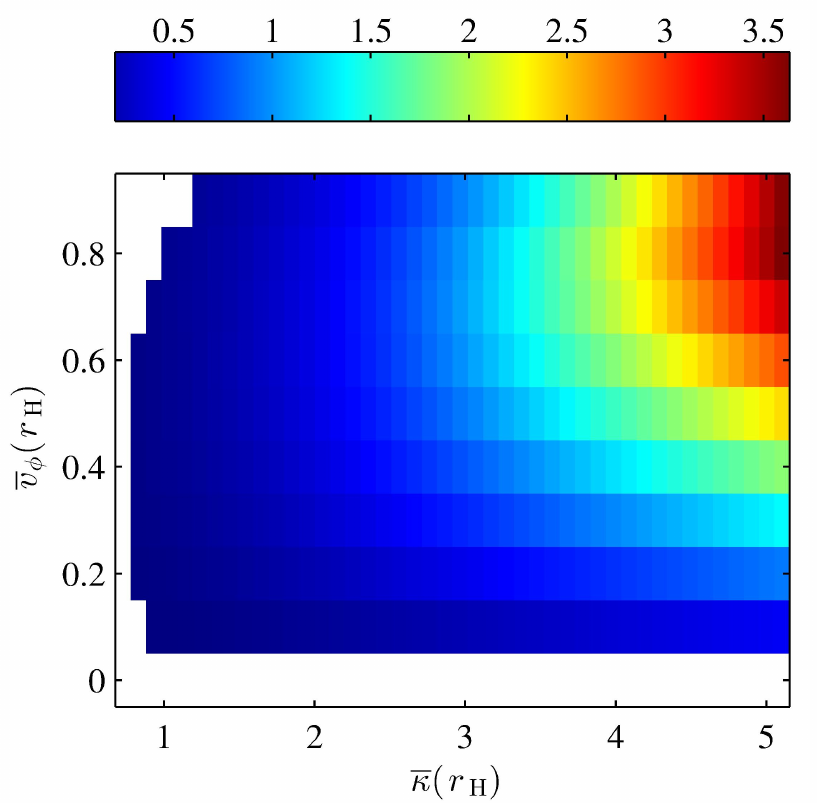}}
\caption{(Colors online.) The left plot shows the maximum possible amplification as a function of $\overline{v}_\phi(r_\mathrm{H})$ and $\overline{\kappa}(r_\mathrm{H})$. The corresponding frequency $\overline{\omega}_\text{peak}$ at which the maximum is attained is shown in the right plot.}\label{F:fig2}%
\end{figure*}

 The scattering problem, on the other hand, is characterized not only by the background parameters $\overline{v}_\phi(r_\mathrm{H})$ and $\overline{\kappa}(r_\mathrm{H})$, but also by the wave parameters $m$ and $\overline{\omega}$. When $m=0$ we conclude from \eqref{supcon} that superradiance is not possible. For non-zero $m$, we have verified that the effect is maximized when $m=1$, similarly to what happens for a Kerr black hole. Consequently, we focus our analysis on the case $m=1$. Following the procedure described in the last section, we solve Eq.~\eqref{radialeqn} with the appropriate boundary conditions and plot the spectrum of reflection coefficients for several parameters $\overline{v}_\phi(r_\mathrm{H})$ and $\overline{\kappa}(r_\mathrm{H})$. The results are shown in Fig.~\ref{F:fig3}. For each curve in the spectrum, corresponding to a pair of parameters $\overline{\kappa}(r_\mathrm{H})$ and $\overline{v}_\phi(r_\mathrm{H})$, we locate the maximum possible reflection coefficient $\mathcal{R}_\text{max}$ and the corresponding frequency, in units of $\omega_\text{disp}$, at which this maximum occurs. The results are plotted in Fig.~\ref{F:fig2}.
 
The analysis of Figs.~\ref{F:fig3} and \ref{F:fig2} is straightforward and provides important information about the possibility of observing superradiant scattering in analogue black hole experiments. First of all we note that, if $\overline{\kappa}(r_\mathrm{H})$ is increased, while $\overline{v}_\phi(r_\mathrm{H})$ is kept constant, the whole spectrum becomes broader and higher, as seen in sub-figure \ref{SpectraConstVnormp4_4cm} for $\overline{v}_\phi(r_\mathrm{H})=0.4$. Consequently, as a result of increasing $\overline{v}_\phi(r_\mathrm{H})$, we also increase both the maximum possible amplification $\mathcal{R}_\text{max}$ and the frequency at which this maximum occurs, as one can observe in Fig.~\ref{F:fig2}. Since there is no natural upperbound for $\overline{\kappa}(r_\mathrm{H})$ in our model as there is for $\overline{v}_\phi(r_\mathrm{H})$, we expect this monotonic behaviour to continue indefinitely.

The dependence of the spectrum on variations of the normalized angular velocity at the horizon looks similar: as $\overline{v}_\phi(r_\mathrm{H})$ is increased, one can observe both the broadening and the growth in height of the spectrum (see sub-figure \ref{SpectraConstS2100_4cm}). However, there seems to be a limit to such behaviour, which occurs near the critical value $\overline{v}_\phi(r_\mathrm{H})=1$. More precisely, as one can observe in sub-figure \ref{RmaxVersusKappaPaper_4cm}, the maximum possible amplification appears to `saturate' at the sub-critical value $\overline{v}_\phi(r_\mathrm{H}) \approx 0.8$, above which $\mathcal{R}_\text{max}$ starts decreasing again.
   
%
%

%
%
   
    Given the analysis above, it is natural to expect that regimes of high $\overline{\kappa}(r_\mathrm{H})$, near the critical angular velocity $\overline{v}_\phi(r_\mathrm{H})$, provide the best setups for detecting and measuring the superradiant scattering in the laboratory. While it is true that this region of the parameter space provides the largest possible amplification, much of it also falls out of the range of validity of our approximations. First of all, as one can see in Fig.~\ref{F:fig1}, large values of the surface gravity $\overline{\kappa}(r_\mathrm{H})$ imply that the slope of the free surface is not sufficiently small in this region, so that the condition $h'(r)^2 \ll 1$ breaks down there. Another important approximation used in our simulations is the shallow water (linear dispersion) approximation $\omega \lesssim \omega_\text{disp}$, which we would like to be satisfied by the peak frequency $\overline{\omega}_\text{peak}$. However, as sub-figure \ref{OmegaMaxOverOmegaDispersionPaper_4cm} shows, for the highest possible amplifications, the peak frequencies sit well outside the linear dispersion regime ($\overline{\omega}<1$). 

Nonetheless, even though we cannot trust our model in the regimes where the largest possible amplifications lies, there is still an interesting region of the parameter space which seems to produce detectable amplifications in the laboratory. Indeed, the left top corner of both plots in Fig.~\ref{F:fig2}, corresponding approximately to $ 1 \lesssim \overline{\kappa}(r_\mathrm{H}) \lesssim 2$ and $ 0.7 \lesssim \overline{v}_\phi(r_\mathrm{H}) \lesssim 0.9$, indicate a regime for which the maximum amplification can be as large as $\mathcal{R}_\text{max} \approx 1.4$, while the peak frequency satisfies $\overline{\omega}_\text{peak} \lesssim 1$. This regime corresponds to the top left corner of Fig.~\ref{F:fig1}, where the slope condition $h'(r)^2 \ll 1$ is satisfied.

We would also like to compare our findings with the results of Refs.~\cite{basak,basak2} for pressure waves. In particular, an analytical expression for the reflection coefficient, valid in the small frequency regime, is obtained in Ref.~\cite{basak2} through a Starobinski-like tecnique~\cite{staro1}. First of all, it is not clear that an analogous calculation for small frequencies will hold in our system since, in our case, the velocities $c_{\mathrm{gw}}$ and $\left. v_r \right|_{z=h}$ are complicated functions of r and, consequently, the location of the horizon can only be determined numerically. Therefore, an analytical comparison in the small frequency regime seems impossible. 
For a numerical comparison, on the other hand, we have to resort to Ref.~\cite{berti}, which provides a numerical implementation of exactly the same equations found in Refs.~\cite{basak,basak2}. This implementation, like ours, covers not only the small frequency regime, but the entire superradiant range. It shows that larger amplification factors occur for larger angular velocity parameters $B$. Our results are similar, although there seems to be a saturation of this effect near the critical velocity, as explained before. Another important difference is that our maximum amplification, for fixed $m$, depends on two parameters, while in the case of Refs.~\cite{basak,basak2,berti}, like for Kerr black holes, it depends on only one, namely $\hat{B}=B/A$. Finally, it is not completely clear in Ref.~\cite{berti} which values of the free parameter are realistic, but for $\hat B=1$ (analogous to our critical condition), they obtain $\mathcal{R}_\text{max} \approx 1.212$. In our analysis, for angular velocities near the critical value, we can attain amplifications of order $\mathcal{R}_\text{max} \approx 1.4$ if we require the surface gravity to be in a region where our approximations hold, as explained before.

Regarding the experimental feasibility of such setups, to the best of our knowledge there has never been an experimental realization of analogue black holes based on the propagation of sound waves in water. Some difficulties associated with sound wave black holes are the possibility of shock waves for fluid velocities approaching the sound speed and the fact that the sound speed ($\approx 1480 m/s$) is typically much larger than the velocity of gravity waves ($\approx \sqrt{gh} $). On the other hand, analogue black hole experiments based on gravity waves are relatively common in the literature, see e.g.~\cite{unruh_3,rousseaux,weinfurtner,weinfurtner2,jump}. Our simulations are the first to attempt a realistic prediction of superradiant amplification in a realistic setup.

\section{Final Remarks} \label{Sec:finalremarks}

In this work we have studied superradiant scattering of shallow water gravity waves impinging on a stationary draining water vortex as a rotating black hole analogue. By using a combination of theory and numerical simulations, we have calculated the reflection coefficients for incident waves and have shown that there exists a window of physical parameter space where our approximations are satisfied and superradiant amplification is predicted. It is important to note that this effect is different from the hydrodynamic analogue of the Aharonov-Bohm effect, which is characterized by dislocation and discontinuity in the incident wavefronts~\cite{ab1,ab2,ab3,ab4}.

The reduction of the dimension of the parameter space, described in Sec.~\ref{Sec:rescale}, is crucial for understanding our numerical results and comparing our system with the superradiant amplification by Kerr black holes. In the general relativistic case, the only background parameter relevant for the scattering process is $a/M$, while in our analogue black hole system there are two important background parameters, $\overline{v}_\phi(r_\mathrm{H})$ and $\overline{\kappa}(r_\mathrm{H})$. From a purely mathematical point of view there is nothing special in these two parameters. One could have used instead the pair ($\overline{A}$, $\overline{B}$), or any other combination of them, in order to analyze the problem. From a physical point of view, however, the choice of $\overline{v}_\phi(r_\mathrm{H})$ and $\overline{\kappa}(r_\mathrm{H})$ is natural. First of all, the parameter $\overline{v}_\phi(r_\mathrm{H})$ is the equivalent of $a/M$ for Kerr black holes and both have an upper limit of $1$. For Kerr black holes, the maximum possible amplification is attained for almost exactly extremal configurations $a/M \approx 1$. For our analogue system, on the other hand, the maximum is attained at a sub-critical value $\overline{v}_\phi(r_\mathrm{H})\approx 0.8$. The choice of $\overline{v}_\phi(r_\mathrm{H})$ therefore provides a natural way to compare our analogue system with Kerr black holes.

The scattering in our analogue black hole, being dependent on two parameters instead of one, is more complex than the Kerr black hole scattering. In our analysis, besides $\overline{v}_\phi(r_\mathrm{H})$, we chose to use the normalized surface gravity $\overline{\kappa}(r_\mathrm{H})$ since it is an important quantity for both real and analogue black holes and plays an important role in other phenomena, for example Hawking radiation. Our simulations show that the maximum amplification increases monotonically as a function of $\overline{\kappa}(r_\mathrm{H})$, suggesting that this parameter is indeed an important controlling parameter for the superradiant amplification.  

In summary, we believe that our analysis uncovers convincing evidence that a draining water 
vortex flow is sufficiently complex and rich to reproduce and generalize many interesting 
features of Kerr black hole superradiance . Given our numerical analysis we also conclude that 
experimental observation of this phenomenon is within reach of forthcoming experiments.

\acknowledgments{We thank Andrew Jason Penner for his help during the initial stages of this project. We also thank Thomas Sotiriou for enlightening discussions on astrophysical rotating black holes within the framework of general relativity and beyond, and Carlos Herdeiro and Vitor Cardoso for enlightening discussions about the maximum amplification for Kerr black holes. MR would like to acknowledge financial support from the S\~ao Paulo Research Foundation (FAPESP), grants \#2013/09357-9 and \#2013/15748-0. SW was funded by a Royal Society University Research Fellowship (URF), a Nottingham Research Fellowship (NRF), and a Royal Society Project Grant. S.L. acknowledges financial support from the John Templeton Foundation (JTF), grant \#51876.}


\bibliographystyle{apsrev}
\bibliography{super_njp}

\end{document}